\documentclass[journal]{IEEEtran}
\usepackage{amsmath,amsfonts}
\usepackage{algorithmic}
\usepackage{algorithm}
\usepackage{array}
\usepackage[caption=false,font=normalsize,labelfont=sf,textfont=sf]{subfig}
\usepackage{textcomp}
\usepackage{stfloats}
\usepackage{url}
\usepackage{verbatim}
\usepackage{graphicx}
\usepackage{booktabs}
\usepackage{enumitem}
\usepackage{tabularx}

\newcolumntype{C}{>{\centering\arraybackslash}X}
\newcolumntype{L}{>{\raggedright\arraybackslash}X}
\newcolumntype{R}{>{\raggedleft\arraybackslash}X}
\newcolumntype{P}[1]{>{\raggedright\arraybackslash}p{#1}}
\usepackage{siunitx}
\usepackage{etoolbox}                           %
\newcommand{\ubold}{\fontseries{b}\selectfont}  %
\robustify\ubold                                %

\newcommand{\tablecaptionsep}{\vspace*{-5pt}}

\usepackage{multirow}
\usepackage{caption}
\usepackage{microtype}
\usepackage{enumitem}
\usepackage{amssymb}
\usepackage{hyperref}

\usepackage{cite}
\def\papertitle{Disentanglement in a GAN for Unconditional Speech Synthesis}
\def\modelname{{ASGAN}}

\usepackage{xcolor}

\begin{document}

\title{\papertitle}

\author{Matthew Baas,~\IEEEmembership{Student Member,~IEEE,} and Herman Kamper,~\IEEEmembership{Senior Member, IEEE}
\thanks{Matthew Baas and Herman Kamper are with the Department of Electrical and Electronic Engineering, Stellenbosch University, South Africa (email: \href{mailto:20786379@sun.ac.za}{20786379@sun.ac.za}; \href{mailto:kamperh@sun.ac.za}{kamperh@sun.ac.za}).} %
\thanks{
All experiments were performed on Stellenbosch University’s High Performance Computing (HPC) GPU cluster.
}
\thanks{Manuscript accepted to IEEE TASLP, 2024.}}

\markboth{%
Accepted to IEEE TASLP, 2024}%
{Baas and Kamper \MakeLowercase{\textit{et al.}}: \papertitle}

\IEEEpubid{\copyright~2024 IEEE}

\maketitle

\begin{abstract}

Can we develop a model that can synthesize realistic speech directly from a latent space, without explicit conditioning?
Despite
several efforts
over the last decade, previous adversarial and diffusion-based approaches still struggle to achieve this, even on small-vocabulary datasets.
To address this, we propose AudioStyleGAN (\modelname) -- a generative adversarial network %
for unconditional speech synthesis tailored to learn a disentangled latent space.
Building upon the StyleGAN family of image synthesis models, \modelname~maps sampled noise to a disentangled latent vector which is then mapped to
a sequence of audio features so that signal aliasing is suppressed at
every layer. 
To successfully train \modelname, we introduce a number of new techniques, including a modification to adaptive discriminator augmentation which probabilistically skips discriminator updates. 
We apply it on the small-vocabulary Google Speech Commands digits dataset, where it achieves state-of-the-art results in unconditional speech synthesis. 
It is also substantially faster than existing top-performing diffusion models.
We confirm that \modelname's latent space is disentangled: we demonstrate how simple linear operations in the space can be used to perform several tasks unseen during training.
Specifically, we perform %
evaluations in voice conversion, speech enhancement, speaker verification, and keyword classification.
Our work indicates that GANs are still highly competitive in the unconditional speech synthesis landscape, and that disentangled latent spaces can be used to aid generalization to unseen tasks.
Code, models, samples:
{\footnotesize \url{https://github.com/RF5/simple-asgan/}}.

\end{abstract}


\section{Introduction} \label{sec:1_intro}

\IEEEPARstart{U}{nconditional} speech synthesis systems aim to produce coherent speech without  conditioning inputs such as text or speaker labels \cite{adv_audio_synth_donahue2018adversarial}.
In this work we are specifically
interested in learning to map noise from a known continuous distribution into %
spoken utterances~\cite{beguvs2020generative}.
A model that could do this would have several useful downstream applications:
from latent interpolations between utterances and fine-grained tuning of properties of the generated speech, to audio compression and
better probability density estimation of speech.
Some of these advances from latent generative modelling have already been realized in the image modality~\cite{stylegan3_karras2021alias,pan2023_DragGAN}; our goal is to bring these developments to the speech domain.

Direct speech synthesis from a latent space is a very challenging problem, 
with prior efforts only able to model speech in a restricted setting~\cite{adv_audio_synth_donahue2018adversarial,diffwave_kong2020,sashimi_goel2022s}.
We therefore also restrict ourselves to the small-vocabulary case here.

In this problem setting, recent studies
on diffusion models~\cite{diffusion_sohl2015deep} for images \cite{dall-e_ramesh2021zero,dall-e-2_ramesh2022hierarchical,imagen_saharia2022photorealistic} has led to major improvements in unconditional speech synthesis. %
The current best-performing approaches are all based on diffusion modelling \cite{sashimi_goel2022s, diffwave_kong2020}, which iteratively de-noise a sampled signal into a waveform through a Markov chain~\cite{diffusion_sohl2015deep}.
Before this, many studies used generative adversarial networks~(GANs) \cite{gans_goodfellow2014generative} 
that map a latent vector to a sequence of speech features with a single forward pass through the model.
However, performance was limited  \cite{adv_audio_synth_donahue2018adversarial, beguvs2020generative}, leading to GANs falling out of favour for this task.

Motivated by the recent developments in the StyleGAN literature \cite{stylegan1_karras2019style,stylegan2_karras2020analyzing,stylegan3_karras2021alias} for image synthesis, we aim to reinvigorate GANs for unconditional speech synthesis, where we are particularly interested in their ability for learning continuous, disentangled latent spaces \cite{stylegan2_karras2020analyzing}.
To this end, we propose AudioStyleGAN~(\modelname{}): a convolutional
GAN which maps a single latent vector to a sequence of audio features, and is designed to have a disentangled latent space.
The model is based in large part on StyleGAN3 \cite{stylegan3_karras2021alias}, which we adapt for audio synthesis. 
Concretely, we adapt the style layers to remove signal aliasing caused by the non-linearities in the network. 
This is accomplished with anti-aliasing filters to ensure that the Nyquist-Shannon sampling limits are met in each layer. 
We also propose a modification to adaptive discriminator augmentation \cite{ada_karras2020training} to stabilize training by randomly dropping discriminator updates based on a guiding signal.

\IEEEpubidadjcol

In unconditional speech synthesis experiments, we measure the quality and diversity of the
generated samples using objective metrics. %
We show that \modelname{} sets a new state-of-the-art in unconditional speech synthesis on the Google Speech
Commands digits dataset~\cite{speechcommands_warden2018speech}. 
Objective metrics that measure latent space disentanglement indicate that \modelname{} has %
a more disentangled latent representation compared to existing diffusion models.
It not only outperforms the best existing models but is also faster to train and faster in inference.
Subjective mean opinion scores (MOS) indicate that \modelname{}’s generated
utterances sound more natural (MOS: 3.68) than the existing best
model (SaShiMi~\cite{sashimi_goel2022s}, MOS: 3.33).
We also perform ablation experiments showing intrinsically that our proposed anti-aliasing and adaptive discriminator augmentation techniques are necessary for high-quality and diverse synthesis.

This work is an extension of the conference paper~\cite{gan_you_hear_me_slt}, where (apart from the ablation experiments) many of the above intrinsic evaluations were already presented. 
Here, for the first time, we 
profile the generalization benefits of a disentangled latent space.
Through an evaluation of \modelname{}'s abilities we
show that its disentangled latent space allows us to perform several tasks unseen during training through simple linear operations in its latent space.
Concretely, we demonstrate compelling zero-shot performance on voice conversion, speech enhancement, speaker verification and keyword classification on the Google Speech Commands digits dataset. %
While not matching the performance of state-of-the-art task-specific systems on all these tasks, our experiments show that a single model designed for disentanglement can achieve reasonable performance across a range of tasks that it hasn't seen in training.
Our work shows the continued competitive nature of GANs compared to diffusion models, and the generalization benefits of designing for disentanglement.

The paper is organized as follows.
We discuss related work in Sec.~\ref{sec:2_related} and then go on to propose \modelname{} in Sec.~\ref{sec:3_model}.
The main unconditional speech synthesis experiments and their results are given in Sec.~\ref{sec:5_exp-setup} and~\ref{sec:6_results}.
This is followed by the experiments on the unseen tasks that \modelname{} can be used for in Sec.~\ref{sec:4_latent_ops}, \ref{sec:exp_setup_unseen}, and~\ref{sec:8_unseen}.

\section{Related Work} \label{sec:2_related}

Since we focus on the proposed generalization abilities provided by a continuous latent space, we first distinguish what we call \textit{unconditional speech synthesis} to the related but different task of \textit{generative spoken language modelling}~(GSLM)~\cite{textless_nlp1_lakhotia2021generative}.
In GSLM, a large autoregressive language model is typically trained on discovered discrete units (e.g.\ HuBERT~\cite{hubert2021} clusters or clustered spectrogram features), similar to how a language model is trained on text~\cite{textless_nlp2_polyak2021speech,textless_nlp3_kharitonov-etal-2022-text}.
While this also enables the generation of speech without any conditioning input, GSLM implies a model structure consisting of an encoder to discretize speech, a language model, and a decoder to convert the discrete units back into a waveform \cite{textless_nlp1_lakhotia2021generative}.
By using discrete units, GSLM approaches can produce
long %
sequences while retaining good performance.
The downside of this discrete approach is that, during generation, you are bound by the discrete units in the model. E.g.\ it is not possible to interpolate between two utterances in a latent space or to directly control speaker characteristics during generation.
If this is desired, additional components must be explicitly built into the model~\cite{textless_nlp3_kharitonov-etal-2022-text}.

In contrast, in unconditional speech synthesis we do not assume any knowledge of particular aspects of speech beforehand. 
Instead of using some intermediate discretization step, such models %
use noise to directly generate speech, often via some latent representation.
The latent space should ideally be disentangled, allowing for better control of the generated speech.
In contrast to GSLM, the synthesis model should learn to disentangle without being explicitly designed to control specific speech characteristics.
In some sense this is a more challenging task than GSLM, which is why most unconditional speech synthesis models are still evaluated on short utterances of isolated spoken words~\cite{adv_audio_synth_donahue2018adversarial} (as we also do here).
In more structured conditional synthesis tasks such as text-to-audio or text-to-video, recent studies~\cite{liu2023audioldm, blattmann2023align_videoldm} have demonstrated the %
benefits of %
modelling a continuous latent space from noise, and then performing synthesis from that latent space.
We aim to apply this reasoning to the unconditional speech synthesis domain in an attempt to realize similar benefits. 

Within unconditional speech synthesis, a substantial body of work focuses on either autoregressive models \cite{oord2016wavenet} -- generating a current sample based on previous outputs -- or diffusion models~\cite{diffwave_kong2020}.
Diffusion models iteratively denoise a sampled signal into a waveform through a Markov chain %
according to a specified inference schedule~\cite{diffusion_sohl2015deep}.
At each inference step, the original noise signal is slightly denoised until, in the last step, it resembles coherent speech.
Autoregressive and diffusion models are relatively slow because they require repeated forward passes through the model during inference.

Earlier studies~\cite{adv_audio_synth_donahue2018adversarial,beguvs2020generative} attempted to use GANs \cite{gans_goodfellow2014generative} for unconditional speech synthesis, which has the advantage of requiring only a single pass through the model.
While results showed some initial promise, performance was poor in terms of speech quality and diversity, with the more recent diffusion models performing much better~\cite{sashimi_goel2022s}.
However, there have been substantial improvements in GAN-based modelling for image synthesis in the intervening years~\cite{ada_karras2020training, stylegan1_karras2019style, stylegan2_karras2020analyzing}.

Palkama et. al.~\cite{palkama20_conditional_speech_stylegan} made initial inroads in applying techniques from the earlier StyleGAN models for unconditional speech synthesis.
Their main focus, however, was to improve conditional speech synthesis using the digit label to guide generation, with the ultimate goal of building a GAN-based text-to-speech model.
Our focus instead is on unconditional speech synthesis, particularly in how a GAN-based approach leads to a disentangled latent space.

In a very related research direction, Beguš et al.~\cite{BEGUS2022101244, beguvs2020generative, beguvs2023basic} and Chen and Elsner~\cite{chen2023exploring} have been studying how GAN-based unconditional speech synthesis models internally perform lexical and phonological learning, and how this relates to human learning.
Most of these studies, however, rely on older GAN synthesis models.
We hope that by developing better performing GANs for unconditional speech synthesis, we can contribute to improving such investigations.
Recently, \cite{audio_stylegan2_2022} attempted to directly use StyleGAN2 for conditional and unconditional synthesis of emotional vocal bursts.
This further motivates a reinvestigation of GANs, but here we look specifically at the generation of speech rather than paralinguistic sounds.

\section{Audio Style GAN} \label{sec:3_model}

\begin{figure*}[t!]
\centering
\centerline{\includegraphics[width=1.0\linewidth]{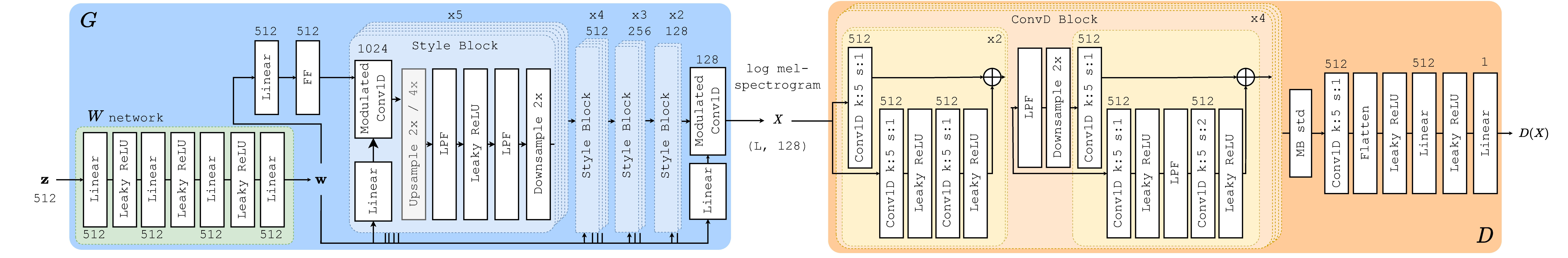}}
\caption{
    The \modelname{} generator $G$ (left) and discriminator $D$ (right). 
    FF, LPF, Conv1D indicate Fourier feature \cite{stylegan3_karras2021alias}, low-pass filter, and 1D convolution layers, respectively.
    The number of output features/channels
    are indicated above linear and convolutional layers.
    Stacked blocks indicate a layer repeated sequentially.
}
\label{fig:1_model_arch}
\end{figure*}

Our model  is based on the StyleGAN family of models~\cite{stylegan1_karras2019style} for image
synthesis. 
We adapt and extend the approach to audio, and therefore
dub our model AudioStyleGAN (\modelname{}).

The model follows the setup of a standard GAN with a single generator network $G$ and discriminator network $D$ \cite{gans_goodfellow2014generative}. 
The generator $G$ accepts a vector $\mathbf{z}$ sampled from a normal distribution and processes it into a sequence of speech features $X$. 
In this work, we restrict the sequence of speech features $X$ to always have a fixed pre-specified duration. 
The discriminator $D$ accepts a sequence of speech features
$X$ and yields a scalar output. 
The model uses a non-saturating logistic loss~\cite{gans_goodfellow2014generative} whereby $D$ is optimized to raise its output for $X$ sampled from real data and minimize its output for $X$ produced by the generator. 
Meanwhile, the generator $G$ is optimized to maximize $D(X)$ for $X$ sampled from the generator, i.e.\ when $X=G(\mathbf{z})$. 
The speech features $X$ are converted to a waveform using a pretrained HiFiGAN
vocoder~\cite{hifi-gan}.
During training, a new adaptive discriminator updating technique is added to ensure stability and convergence.
Each component is described in detail below.

\subsection{Generator} 
The architecture of the generator $G$ is shown on the left of Fig.~\ref{fig:1_model_arch}. 
It consists of a latent mapping network $W$ that converts $\mathbf{z}$ to a disentangled latent space, a special Fourier feature
layer which converts a
single vector from this latent space into a sequence of cosine features
of fixed length, and finally a convolutional encoder which iteratively
refines the cosine features into the final speech features $X$.

\subsubsection{Mapping network}
The mapping network $W$ is a simple multi-layer perceptron consisting of several linear layers with leaky ReLU
activations between them. 
As input, it takes in a normally distributed vector $\mathbf{z} \sim Z = \mathcal{N}(\mathbf{0}, \mathbf{I})$; 
in all our experiments we use a 512-dimensional multivariate normal vector, $\mathbf{z} \in \mathbb{R}^{512}$. 
Passing this vector through the mapping network produces a latent vector $\mathbf{w} = W(\mathbf{z})$ of the same dimensionality as $\mathbf{z}$. 
As explained in~\cite{stylegan1_karras2019style}, the primary purpose of $W$
is learn to map noise to a linearly disentangled $W$ space, as ultimately this will allow for more controllable and understandable synthesis.
$W$ is coaxed into learning such a disentangled representation because
it can only linearly modulate channels of the cosine features in each
layer of the convolutional encoder (see details below). 
So, if $\mathbf{w}$ is to linearly shape the speech features, $W$ %
should ideally
learn a mapping that organizes the random normal space $\mathbf{z}$ into one which linearly disentangles common factors of speech variation.

\subsubsection{Convolutional encoder} \label{subsec:conv_encoder}
The convolutional encoder begins by linearly projecting $\mathbf{w}$ as the input to a 
Fourier feature layer~\cite{fourierfeat_tancik2020} as shown in
Fig.~\ref{fig:1_model_arch}. 
Concretely, we use the Gaussian Fourier feature mapping
from~\cite{fourierfeat_tancik2020} and incorporate the transformation proposed in StyleGAN3~\cite{stylegan3_karras2021alias}. 
This
layer samples a frequency and phase from a
Gaussian distribution for each output channel (fixed at initialization).
The layer then linearly projects the input vector to a vector of phases
which are added to the initial random phases. 
The output sequence is obtained as the plot of a cosine function with these frequencies and phases, one frequency/phase for each output channel. 
Intuitively, the layer provides a way to learn a mapping between a single vector ($\mathbf{w}$) into a \textit{sequence} of vectors at the output of the Fourier feature layer, which provides the base from which the rest of the model can iteratively operate on the feature sequence to eventually arrive at a predicted mel-spectrogram.

The sequence produced by the Fourier feature layer is iteratively passed through 
14 \texttt{Style Blocks}, which are based on the encoder layers of the StyleGAN family of models. 
In each layer, the input sequence and style vector $\mathbf{w}$ are passed through a modulated convolution layer~\cite{stylegan2_karras2020analyzing}:
the final convolution kernel is computed by multiplying the layer’s
learnt kernel with the style vector derived from $\mathbf{w}$, broadcasted over
the length of the kernel. 
In this way, the latent vector $\mathbf{w}$ linearly modulates the kernel in each convolution.

To ensure the signal does not experience aliasing due to the non-linearity, the leaky ReLU layers are surrounded by layers responsible for anti-aliasing (explained below).
All these layers comprise a single \texttt{Style Block}, which is repeated in groups of 5, 4, 3, and finally 2 blocks.
The last block in each group upsamples by $4\times$ instead of $2\times$, thereby increasing the sequence length by a factor of 2 for each group.
A final 1D convolution layer projects the output from the last group into the audio feature space (e.g.\ log mel-spectrogram 
or HuBERT features), as illustrated in the middle of Fig.~\ref{fig:1_model_arch}.

\subsubsection{Anti-aliasing filters}\label{subsec:3_aa_filters}
From image synthesis with GANs~\cite{stylegan3_karras2021alias}, we know that the generator must include anti-aliasing filters for the signal propagating through the network to approximately satisfy the Nyquist-Shannon sampling theorem.
This is why, before and after a non-linearity, we include upsampling, low-pass filter (LPF), and downsampling layers in each \texttt{Style Block}.
The motivation from \cite{stylegan3_karras2021alias} is that 
non-linearities introduce arbitrarily high-frequency information into the output signal.
The signal we are modelling (speech) is continuous,
and the internal discrete-time features %
that are passed through the network is therefore a digital representation of this continuous~signal.
From the Nyquist-Shannon sampling theorem, we know that for such a discrete-time signal to accurately reconstruct the continuous signal, it must be bandlimited to $\SI{0.5}{\text{cycles/sample}}$.
If not, \cite{stylegan3_karras2021alias} showed that the generator learns to use aliasing artefacts to fool the discriminator, to the detriment of quality and control of the final output.
To address this, we follow \cite{stylegan3_karras2021alias}:
we approximate an ideal continuous LPF 
by first upsampling to a higher sample rate, applying a discrete LPF as a 1D convolution, and only then applying the non-linearity. 
This high-frequency signal is then passed through an anti-aliasing discrete LPF before being downsampled again to the original sampling rate.

Because of the practical imperfections in this anti-aliasing
scheme, the cutoff frequencies for the LPFs must typically be much
lower than the Nyquist frequency of $\SI{0.5}{\text{cycles/sample}}$.
We reason that the generator should ideally first focus on generating coarse features before generating good high-frequency details, which will inevitably
contain more trace aliasing artifacts. So we design the filter cutoff
to begin at a small value in the first \texttt{Style Block}, and increase gradually to near the critical Nyquist frequency in the final block.
In this way, aliasing is kept fairly low throughout the network,
with very high frequency information near the Nyquist frequency
only being introduced in the last few layers.

\subsection{Discriminator}

The discriminator $D$ has a convolutional architecture similar to~\cite{stylegan2_karras2020analyzing}.
It takes a sequence of speech features $X$ as input and predicts whether it is generated by $G$ or sampled from the dataset.
Concretely, $D$ consists of four \texttt{ConvD Blocks} and a network head, as show in Fig.~\ref{fig:1_model_arch}.
Each \texttt{ConvD Block} consists of 1D convolutions with skip connections, and a downsampling layer with an anti-aliasing LPF in the last skip connection.
The LPF cutoff is set to the Nyquist frequency for all layers.
The number of layers and channels are chosen so that $D$ has roughly the same number of parameters as $G$.
$D$'s head consists of a minibatch standard deviation~\cite{ema_gen_weights_karras2018progressive} layer and a 1D convolution layer before passing the flattened activations through a final linear projection head to arrive at the logits.
Both $D$ and $G$ are trained using the non-saturating logistic loss \cite{gans_goodfellow2014generative}.

\subsection{Vocoder}

The generator $G$ and discriminator $D$ operate on sequences of speech
features and not on raw waveform samples. 
Once both $G$ and $D$ are trained, we need a way at convert these speech features back to waveforms. 
For this we use a pretrained HiFi-GAN vocoder~\cite{hifi-gan} that vocodes either log mel-scale spectrograms~\cite{mel-scale} or HuBERT features~\cite{hubert2021}.
HuBERT is a self-supervised speech representation model that learns to encode speech in a 50~Hz vector sequence using %
a masked token prediction task.
These learnt features are linearly predictive of several high-level characteristics of speech such as phone identity, making it useful as a feature extractor when trying to learn disentangled representations.

\subsection{Implementation} \label{subsec:3_model_implementation}

We train two variants of our model: a log mel-spectrogram
model and a HuBERT model.
And, extending our initial investigation in~\cite{gan_you_hear_me_slt}, we train additional variants of the HuBERT-based model to understand key design choices.

\subsubsection{\modelname{} variants} 
The log mel-spectrogram model architecture is shown in Fig.~\ref{fig:1_model_arch}, where mel-spectrograms are computed with 128 mel-frequency bins at a hop and window size of 10~ms and 64~ms, respectively. 
Each 10 ms frame of the mel-scale spectrograms is scaled by taking the natural
logarithm of the spectrogram magnitude.
The HuBERT-based model is identical except that it only uses only half the sequence length %
(since HuBERT features are 20~ms instead of the 10~ms spectrogram frames) and has a different number of output channels in the four groups of \texttt{Style Block}s: \texttt{[1024, 768, 512, 512]} convolution channels instead of the
\texttt{[1024, 512, 256, 128]} used for the mel-spectrogram model.
This change causes the HuBERT variant to contain 51M parameters,
as opposed to the 38M parameters in %
the mel-spectrogram
model.

\subsubsection{Vocoder variants} 
The HiFi-GAN vocoder for both HuBERT and mel-spectrogram features is based on the original author’s implementation~\cite{hifi-gan}. 
The HuBERT feature HiFi-GAN is trained 
on the LibriSpeech \texttt{train-lean-100} %
multispeaker speech dataset~\cite{panayotov2015librispeech}
to vocode activations extracted from layer 6 of the pretrained HuBERT \texttt{Base} model provided with
fairseq~\cite{ott2019fairseq}.
The mel-spectrogram HiFiGAN is trained on the Google Speech Commands dataset.
Both are trained using the original V1 HiFi-GAN configuration (number of updates, learning and batch size parameters) from~\cite{hifi-gan}.

\subsubsection{Optimization} 
Both \modelname{} variants are trained with Adam \cite{adam} ($\beta_1 = 0, \beta_2 = 0.99$), clipping gradient norms at 10, and a learning rate of $3\cdot10^{-3}$ for 520k iterations with a batch size of 32. 
Several critical tricks are used to stabilize GAN training:
(i) equalized learning rate is used for all trainable parameters~\cite{ema_gen_weights_karras2018progressive}; 
(ii) leaky ReLU activations with $\alpha=0.1$; 
(iii) exponential moving averaging for the generator weights (for use during evaluation) \cite{ema_gen_weights_karras2018progressive};
(iv) $R_1$ regularization \cite{r1_reg_mescheder2018training}; and 
(v) a 0.01-times smaller learning rate for the mapping network $W$,
since it needs to be updated slower compared to the convolution layers in the main network branch %
\cite{stylegan3_karras2021alias}.

\subsubsection{Adaptive discriminator updates} \label{subsubsec:ada_updates}
We also introduce a new technique for updating the discriminator.
Concretely, we first scale $D$'s learning rate by 0.1 compared to the generator as otherwise we find it overwhelms $G$ early on in training.
Additionally we employ a dynamic method for updating $D$, inspired by adaptive discriminator augmentation \cite{ada_karras2020training}:
during each iteration, we skip $D$'s update with probability $p$.
The probability $p$ is initialized at 0.1 and is updated every 16th generator step or whenever the discriminator is updated.
We keep a running average $r_t$ of the proportion of $D$'s outputs on real data $D(X)$ that are positive (i.e. that $D$ can confidently identify as real).
Then, if $r_t$ is greater than 0.6 we increase $p$ by 0.05 (capped at 1.0), and if $r_t$ is less than 0.6 we decrease $p$ by 0.05 (limited %
at 0.0).
In this way, we adaptively skip discriminator updates. 
When $D$ becomes too strong, both $r_t$ and $p$ rise, and so $D$ is updated less frequently. 
Conversely, when $D$ becomes too weak, it is updated more frequently.
We found this new modification to be critical for ensuring that $D$ does not overwhelm $G$ during training.

We also use the traditional adaptive discriminator augmentation \cite{ada_karras2020training} 
where we apply the following transforms to the model input with the same probability $p$:
(i) adding Gaussian noise with $\sigma = 0.05$; (ii) random scaling by a factor of $1 \pm 0.05$; and (iii) randomly replacing a subsequence of frames from the generated speech features with a subsequence of frames taken from a real speech feature sequence.
This last augmentation is based on the fake-as-real GAN method \cite{fake-as-real_tao2020alleviation} and is important to prevent gradient explosion later in training.

\subsubsection{Anti-aliasing filters} \label{subsec:anti_alias_filter}
For the anti-aliasing LPF filters we use windowed \texttt{sinc} filters with a width-9 Kaiser window \cite{kaiser1966digital}.
For the generator (all variants), the first \texttt{Style Block} has a cutoff at $f_c = 0.125\ \text{cycles/sample}$ which is increased in an even logarithmic scale to $f_c=0.45\ \text{cycles/sample}$ in the second-to-last layer, keeping this value for the last two layers to fill in the last high frequency detail.
Even in these last layers we use a cutoff below the Nyquist frequency to ensure the imperfect LPF still sufficiently suppresses aliased frequencies. 
For the discriminator we are less concerned about aliasing as it does not generate a continuous signal, so we
use a $f_c=0.5\ \text{cycles/sample}$ cutoff for all \texttt{ConvD Block}s.

All models are trained using mixed FP16/FP32 precision on a single NVIDIA Quadro RTX 6000 using PyTorch 1.11.
Trained models and code are available at {\url{https://github.com/RF5/simple-asgan/}}.

\section{Experimental Setup: Unconditional Speech Synthesis} \label{sec:5_exp-setup}

\subsection{Data}
To compare to existing unconditional speech synthesis models, %
we use the Google Speech Commands dataset of isolated spoken words \cite{speechcommands_warden2018speech}.
As in other studies \cite{sashimi_goel2022s, diffwave_kong2020, adv_audio_synth_donahue2018adversarial}, we use the
subset corresponding to the ten spoken digits ``zero'' to ``nine'' (called SC09).
The digits are spoken by various speakers under different channel conditions.
This makes it a challenging benchmark for unconditional speech synthesis.
All utterances are roughly a second long and are sampled at 16~kHz, where utterances less than a second are padded to a full second. 

\subsection{Evaluation metrics}\label{subsec:5-uncond-metrics}

We train and validate our models on the official training/validation/test split from SC09. 
We then evaluate unconditional speech synthesis quality by seeing how well newly generated utterances match the distribution of the SC09 test split.
We use metrics similar to those for image synthesis; they try to measure either the \textit{quality} of generated utterances (realism compared to actual audio in the test set), %
or the \textit{diversity} of generated utterances (how varied the utterances are relative to the test set), or a combination of both.

These metrics require extracting features or predictions from a supervised speech classifier network trained to classify the utterances from SC09 by what digit is spoken.
While there is no consistent pretrained classifier used for this purpose, we opt
to use a ResNeXT architecture \cite{resnext_xie2017aggregated}, similar to previous studies \cite{sashimi_goel2022s, diffwave_kong2020}.
The trained model has a 98.1\% word classification accuracy on the SC09 test set, and we make the model code and checkpoints available for future comparisons.\footnote{\scriptsize\url{https://github.com/RF5/simple-speech-commands}} 
Using either the classification output or 1024-dimensional features extracted from the penultimate layer in the classifier,
we consider the following metrics.

\begin{itemize}
    \item \textit{Inception score} (IS) measures the diversity and quality of generated samples by evaluating the Kullback-Leibler (KL) divergence between the label distribution from the classifier output and the mean label distribution over a set of generated utterances~\cite{is_salimans2016improved}.
    \item \textit{Modified Inception score} (mIS) extends the diversity measurement aspect of IS by incorporating a measure of intra-class diversity (in our case over the ten digits) to reward models with higher intra-class entropy \cite{modified_is_gurumurthy2017deligan}.
    \item \textit{Fréchet Inception distance} (FID) computes a measure of how well the distribution of generated utterances matches the test-set utterances by comparing the classifier features of generated and real data \cite{fid_heusel2017gans}.
    \item \textit{Activation maximization} (AM) measures generator quality by comparing the KL divergence between the classifier class probabilities from real and generated data, while penalizing high classifier entropy samples produced by the generator \cite{am_zhou2018activation}. 
    Intuitively, this attempts to account for possible class imbalances in the training set and intra-class diversity by incorporating a term for the entropy of the classifier outputs for generated samples.
\end{itemize}

A major motivation for \modelname{}'s design is latent-space disentanglement.
In the experiments proceeding from Sec.~\ref{sec:8_unseen} and onwards, we show this property allows the model to be applied to extrinsic tasks that it is not explicitly trained for. But before this (Sec.~\ref{sec:6_results}), we intrinsically evaluate disentanglement using two
metrics on the $Z$ and $W$ latent spaces.
\begin{itemize}
\item 
\textit{Path length} measures the mean $L_2$ distance moved by the classifier features when the latent point ($\mathbf{z}$ or $\mathbf{w}$) is randomly perturbed slightly, averaged over many perturbations \cite{stylegan1_karras2019style}.
A lower value indicates a smoother latent space.
\item
\textit{Linear separability} %
utilizes a linear support vector machine (SVM) to classify the digit of a latent point.
The metric is computed as the additional information (in terms of mean entropy) necessary to correctly classify an utterance %
given the class prediction from the linear SVM~\cite{stylegan1_karras2019style}.
A lower value indicates a more linearly disentangled latent space.
\end{itemize}
These metrics are averaged over 5000 generated utterances for each model.
As in \cite{stylegan1_karras2019style}, for linear separability we exclude half the generated utterances for which the ResNeXT classifier is least confident in its prediction.

To give an indication of naturalness, we compute an estimated mean opinion score (eMOS) using a pretrained \texttt{Wav2Vec2 small} baseline from the VoiceMOS challenge~\cite{huang2022voicemos}.
This model is trained to predict the naturalness score that a human would assign to an utterance from 1 (least natural) to 5 (most natural).
We also perform an actual subjective MOS evaluation using %
Amazon Mechanical Turk to obtain 240 opinion scores for each model with 12 speakers listening to each utterance.
Finally, the speed of each model is also evaluated to highlight
the benefit that GANs can produce
utterances in a single inference call, as opposed to the many inference calls necessary %
with autoregressive or diffusion models.

\subsection{Baseline systems}

We compare to the following unconditional speech synthesis methods (Sec.~\ref{sec:2_related}):
WaveGAN \cite{adv_audio_synth_donahue2018adversarial}, DiffWave \cite{diffwave_kong2020}, autoregressive SaShiMi and SaShiMi+DiffWave~\cite{sashimi_goel2022s}.
Last-mentioned is the current best-performing model on SC09.
For WaveGAN we use the trained model provided by the authors \cite{adv_audio_synth_donahue2018adversarial}, while for DiffWave we use an open-source pretrained model.\footnote{\scriptsize
\url{https://github.com/RF5/DiffWave-unconditional}}
For the autoregressive SaShiMi model, we use the code provided by the authors to train an unconditional SaShiMi model on SC09 for 1.1M updates \cite{sashimi_goel2022s}.\footnote{\scriptsize
\url{https://github.com/RF5/simple-sashimi}
}
Finally, for the SaShiMi+DiffWave diffusion model, we modify the autoregressive SaShiMi code and combine it with DiffWave according to~\cite{sashimi_goel2022s}; we train it on SC09 for 800k updates with the hyperparameters in the original paper \cite{sashimi_goel2022s}.\footnotemark[3]

\modelname{} is a two-stage model (it generates a feature representation, then vocodes the result), while WaveGAN, SaShiMi, and DiffWave are one-stage models (they directly generate the output waveform).
Despite SaShiMi+DiffWave being the current top-performing model, our two-stage approach may have an advantage over one-stage approaches because it separates the vocoding task to a separate model.

Lastly, the autoregressive SaShiMi is originally evaluated using a form of rejection sampling \cite{sashimi_goel2022s} to retain only high probability generated samples for evaluation. %
But, when sampling from a GAN using $\mathbf{z}\sim \mathcal{N}(\mathbf{0}, \mathbf{I})$, we also have the exact density value for this sampled $\mathbf{z}$.
So, we could perform various sampling tricks on all the models (since both diffusion and GAN models also have tractable likelihood measures). However, in the interest of an fair %
and simple comparison of the inherent performance of each model, we opt to keep the same and most general sampling method for each model (including \modelname{}).
So, in all experiments, we perform direct sampling from the latent space for the GAN and diffusion models according to the original papers.
For the autoregressive models, we directly sample from the predicted distribution for each time-step. %

\section{Results: Unconditional Speech Synthesis} \label{sec:6_results}

\subsection{Comparison to baselines}

\setlength{\tabcolsep}{3.6pt}
\begin{table}[!tb]
    \renewcommand{\arraystretch}{1.2}
    \centering
    \caption{
        Results measuring the quality and diversity of generated samples from unconditional speech synthesis models together with train/test set toplines for the SC09 dataset. Subjective MOS values with 95\% confidence intervals are shown.
        IS, mIS, FID, AM are averaged over 5000 generated utterances generated from each model (or drawn from the train/test set for topline scores).
    }
    \tablecaptionsep
    \label{tab:1_quality_diversity}
    
    \begin{tabularx}{1.0\linewidth}{@{}
        L
        S[table-format=1.2]
        S[table-format=3.1]
        S[table-format=1.2]
        S[table-format=1.2]
        S[table-format=1.2]
        S[table-format=1.2(2),
            table-figures-uncertainty=1,
            separate-uncertainty = true]
        @{}}
    \toprule
    Model & {IS$\ \uparrow$} & {mIS $\ \uparrow$} & {FID$\ \downarrow$} & {AM$\ \downarrow$} & {eMOS$\ \uparrow$} & {MOS$\ \uparrow$} \\
    \midrule
    \textit{Train set} & 9.37 & 237.6 & 0 & 0.20 & 2.41 & 3.74(12)\\
    \textit{Test set} & 9.36 & 242.3 & 0.01 & 0.20 & 2.43 & 3.88(12)\\
    \midrule
    WaveGAN \cite{adv_audio_synth_donahue2018adversarial} & 4.45 & 34.6 & 1.77 & 0.81 & 1.06 & 2.88(16) \\
    DiffWave \cite{diffwave_kong2020} & 5.13 & 49.6 & 1.68 & 0.68 & 1.66 & 3.43(14) \\
    SaShiMi \cite{sashimi_goel2022s} & 3.74 & 18.9 & 2.11 & 0.99  & 1.58 & 3.19(15) \\
    SaShiMi+DiffWave & 5.44 & 60.8 & 1.01 & 0.61 & 1.89 & 3.33(12)\\
    \modelname{} (mel-spec.) & 7.02 & 162.8 & 0.56 & 0.36 & 1.76 & 3.51(13)\\
    \modelname{} (HuBERT) & \ubold 7.67 & \ubold 226.7 & \ubold 0.14 & \ubold 0.26 & \ubold 1.99 & \ubold 3.68(13) \\
    \bottomrule
    \end{tabularx}
\end{table}

\setlength{\tabcolsep}{6pt}

\begin{table}[!tb]
    \renewcommand{\arraystretch}{1.2}
    \centering
    \caption{
        Latent-space disentanglement and speed metrics. Speed is measured as the number of samples that can be generated per unit time on a single NVIDIA Quadro RTX 6000 using a batch size of 1, given in ksamples/sec.
        Some models do not have a $W$-space (WaveGAN) or any continuous latent space (SaShiMi).
        } %
    \tablecaptionsep
    \label{tab:2_disentanglement}
	\begin{tabularx}{0.99\linewidth}{@{}Lccccr@{}}
		\toprule
		& \multicolumn{2}{c}{{Path length $\downarrow$}} & \multicolumn{2}{c}{Separability $\downarrow$} &  \\
		\cmidrule(l){2-3}  \cmidrule(l){4-5}
		Model & {$Z$} & {$W$} & {$Z$} & {$W$} & {Speed $\uparrow$}  \\
		\midrule
WaveGAN \cite{adv_audio_synth_donahue2018adversarial} & \ubold 1.03\hphantom{$\cdot \text{10}^{\text{0}}$}  & {---} & 4.86 & {---} & \textbf{2214.71} \\
DiffWave \cite{diffwave_kong2020} & 2.72$\cdot \text{10}^{\text{6}}$ & 7.27$\cdot \text{10}^{\text{5}}$  & 6.09 & 6.58 & 0.83 \\
SaShiMi \cite{sashimi_goel2022s} & {---} & {---} & {---} & {---} & 0.14  \\
SaShiMi+DiffWave & 2.89$\cdot \text{10}^{\text{6}}$ & 1.24$\cdot \text{10}^{\text{6}}$ & 4.07 & 2.34 & 0.47 \\
\modelname{} (mel-spec.) & 6.77$\cdot \text{10}^{\hphantom{\text{1}}}$ & 3.21$\cdot \text{10}^{\hphantom{\text{1}}}$ & 1.81 & 1.01 & 875.45 \\
\modelname{} (HuBERT) & 3.50$\cdot \text{10}^{\hphantom{\text{1}}}$ & \ubold 1.84$\cdot \text{10}^{\hphantom{\text{1}}}$ & \textbf{1.40} & \textbf{1.00} & 816.27 \\
		\bottomrule
	\end{tabularx}
\end{table}

We present our headline results in Table~\ref{tab:1_quality_diversity}, where we compare previous state-of-the-art unconditional speech synthesis approaches to the proposed \modelname{} model.
As a reminder, IS, mIS, FID and AM measure generated speech diversity and quality relative to the test set; eMOS and MOS are measures of generated speech naturalness.
We see that both variants of \modelname{} outperform the other models on most metrics.
The HuBERT variant of \modelname{} in particular performs best across all metrics. 
The improvement of the HuBERT \modelname{} over the mel-spectrogram variant is likely because the high-level HuBERT speech representations make it easier for the model to disentangle common factors of speech variation. 
The previous best unconditional synthesis model, SaShiMi+DiffWave, still outperforms the other baseline models, and it appears to have comparable naturalness (similar eMOS and MOS) to the mel-spectrogram \modelname{} variant.
However, it appears to match the test set more poorly than either \modelname{} variant on the other diversity metrics. 

The latent space disentanglement metrics and generation speed of each model are compared
in Table~\ref{tab:2_disentanglement}.
These results are more mixed, with WaveGAN being the fastest model and the one with the shortest latent path length in the $Z$-space.
However, this is somewhat misleading since WaveGAN's samples have low quality (naturalness, as measured by eMOS/MOS) and poor diversity compared to the other models (Table~\ref{tab:1_quality_diversity}).
This means that WaveGAN's latent space is a poor representation of the true distribution of speech in the SC09 dataset, allowing it to have a very small path length as most paths do not span a diverse set of speech variation. 

In terms of linear separability, \modelname{} again yields substantial improvements over existing models.
The results confirm that  \modelname{} has indeed learned a disentangled latent space -- a primary motivation for the model's design.
Specifically, this shows that the idea from image synthesis of using the latent $\mathbf{w}$ vector to linearly modulate convolution kernels can also be applied to speech.
This level of disentanglement allows \modelname{} to be applied to tasks unseen during training, 
evaluated later in Sec.~\ref{sec:8_unseen}.
Regardless of performance, the speed of all the convolutional GAN models (WaveGAN and \modelname{}) is significantly better than the diffusion and autoregressive models, as reasoned in Sec.~\ref{subsec:5-uncond-metrics}.

\subsection{Ablation experiments} \label{subsec:5.2_ablations}

\setlength{\tabcolsep}{3.6pt}
\begin{table}[!tb]
    \renewcommand{\arraystretch}{1.2}
    \centering
    \caption{
        Ablations of model design choices, comparing quality, diversity, and latent-space disentanglement of several %
        \modelname{} variants.
        } %
    \tablecaptionsep
    \label{tab:3_ablations}
	\begin{tabularx}{0.99\linewidth}{@{}
            L
            S[table-format=3.1]
            S[table-format=1.2]
            S[table-format=1.2]
            S[table-format=3.1]
            S[table-format=3.1]
            @{}}
		\toprule
		& \multicolumn{3}{c}{{Quality \& diversity}} & \multicolumn{2}{c}{Path length $\downarrow$}  \\
		\cmidrule(l){2-4}  \cmidrule(l){5-6}
		Variant & {mIS $\ \uparrow$} & {FID$\ \downarrow$} & {eMOS$\ \uparrow$} & {$Z$} & {$W$}  \\
		\midrule

\modelname{} (HuBERT) & \ubold 226.7 & \ubold 0.14 & \ubold 1.99 & 35.0 & 18.4 \\
\hspace{0.1cm} w/o adaptive $D$ updates & 1.4 & 19.27 & 0.69 & 10.8 & 7.6 \\
\hspace{0.1cm} w/o adaptive $D$ augmentation & 2.3 & 13.72 & 0.68 & \ubold 8.4 & \ubold 5.0 \\
\hspace{0.1cm} w/o anti-aliasing filters & 102.7 & 3.31 & 1.81 & 75.8 & 63.1 \\
\hspace{0.1cm} w/o modulated convolution & 2.4 & 10.96 & 0.63 & 111.4 & 48.5 \\
		\bottomrule
	\end{tabularx}
\end{table}
\setlength{\tabcolsep}{6pt}

While the previous comparisons demonstrated the overall success of \modelname{}'s design, we are still not certain which specific decisions from
Sec.~\ref{sec:3_model} are responsible for its performance. 
So, we perform several ablations of the HuBERT \modelname{} model with specific components removed from the full model.
Concretely, we ablate four key design choices:
we train a variant without adaptive discriminator updates (Sec.~\ref{subsubsec:ada_updates}), a variant without adaptive discriminator augmentation (Sec.~\ref{subsubsec:ada_updates}), %
and
a variant without any anti-aliasing filters 
(Sec.~\ref{subsec:3_aa_filters}).
Finally, we also train a variant without modulated convolutions (Sec.~\ref{subsec:conv_encoder}) such that $\mathbf{w}$ only controls the initial features passed to the generator's convolutional encoder.

Table~\ref{tab:3_ablations} shows the results for the ablated \modelname{} approaches on a subset of the metrics.
We see that on %
the quality and diversity metrics, %
the base \modelname{} is best, while the models without adaptive discriminator updates and augmentation have better latent space disentanglement.
However, recall that the main reason for including these was \textit{not} to optimize disentanglement, but rather to ensure training stability and performance.
As reasoned in Sec.~\ref{sec:3_model}, without either adaptive updates or augmentation, the discriminator has a much easier task and begins to dominate the generator, confidently 
distinguishing between
real 
and
generated samples.
So, while
this makes optimization easier
(leading to a smoother latent space), it means that the generator does not effectively learn from the adversarial task. %
A similar phenomenon can be seen with WaveGAN in Table~\ref{tab:2_disentanglement}, where it 
scored well on the %
disentanglement metrics 
but had poor output quality in Table~\ref{tab:1_quality_diversity}.

When anti-aliasing filters are removed, both latent space disentanglement and synthesis quality are reduced, being slightly worse in all metrics compared to the full model.
This validates our design motivation for the inclusion of low-pass filters to suppress aliased high-frequency content in the layer activations in Fig~\ref{fig:1_model_arch}.
Finally, the variant without the linear influence of $\mathbf{w}$ on each layer's activations (i.e.\ without the modulated convolutions) is also worse than the baseline model in all metrics considered.

Overall, we can see from Table~\ref{tab:3_ablations} that 
each of the key design aspects from Sec.~\ref{sec:3_model} are necessary to achieve both high latent space disentanglement and synthesis quality in a single %
model -- the main requirement for it to be performant on unseen downstream tasks, which we look at next.

\section{Solving Unseen Tasks Through Linear Latent Operations} \label{sec:4_latent_ops}

We have now shown intrinsically that \modelname{} leads to a disentangled latent space. In this section and the ones that follow we show that \modelname{} can also be used extrinsically to perform tasks unseen during training through linear manipulations of its latent space.
From this point onwards we the HuBERT \modelname{} variant.

As a reminder, the
key aspect of \modelname{}'s %
design %
is that the latent space associated with the vector $\mathbf{w}$ is linearly disentangled.
The idea is that, since the
$\mathbf{w}$ vector can only linearly affect the output of the model through the Fourier feature layer and
modulated convolutions (Fig.~\ref{fig:1_model_arch}), %
$W$ should ideally %
learn to linearly disentangle common factors of speech variation.
If this turns out to be true -- i.e.\ the space is indeed disentangled -- then
these factors
should %
correspond to linear %
directions in the $W$ latent space. 
As originally motivated in studies on image synthesis~\cite{stylegan1_karras2019style, stylegan2_karras2020analyzing}, 
this would mean that linear operations
in the latent space
should correspond to meaningful edits in the generated output.
In our case, this would mean that
if we know the %
relationship between 
two
utterances, then the \textit{linear} distance between the latent vectors $\mathbf{w}$ of those utterances should reflect that %
relationship.
E.g. 
consider a collection of %
utterances differing only in noise level but otherwise having the same properties.
If the space disentangles noise levels, then we should
expect the latent points of these utterances to %
lie %
in the same linear subspace, reflective of the level of noise.

\subsection{Projecting to the latent space} \label{subsec:projection}
Before defining how unseen tasks can be phrased as latent operations, we first need to explain how we invert a provided utterance to a latent vector $\mathbf{w}$.
We use a method similar to~\cite{stylegan2_karras2020analyzing} where we optimize a $\mathbf{w}$ vector while keeping $G$ and the  speech feature sequence $X$ fixed. 
Concretely, $\mathbf{w}$ is initialized to the mean $\Bar{\mathbf{w}}= \mathbb{E} [ W(\mathbf{z}) ]$ vector over 100k samples
and then fed through the network to produce a candidate sequence $\Tilde{X}$.
An $L_2$ loss is then formed as the mean square distance between each feature in the candidate sequence $\Tilde{X}$ and the target sequence $X$. 
Optimization follows~\cite{stylegan2_karras2020analyzing}, using Adam for 1000 iterations with a maximum learning rate of 0.1, and with Gaussian noise added to $\mathbf{w}$ in the first 750 iterations. The variance of this noise is set to be %
proportional to the average square $L_2$ distance between the mean $\Bar{\mathbf{w}}$ and the sampled $\mathbf{w}$ vectors. %

\subsection{Downstream tasks}

We look at several downstream tasks, each of which can be phrased as linear operations in the latent $W$-space.

\subsubsection{Style mixing} \label{subsec:style_mix_explanation}
We can perform voice conversion or speech editing by using style mixing of the latent vectors $\mathbf{w}$.
Style mixing is a technique proposed in~\cite{stylegan1_karras2019style}, where they find that StyleGAN models encode coarse information (e.g. facial type, frame positioning) using the $\mathbf{w}$ vector passed to earlier layers, and finer details (e.g. hair color, lighting) using the $\mathbf{w}$ vector passed to later layers.
We will qualitatively show that this also applies to speech in \modelname{}, where earlier layers control coarse aspects of speech (e.g.\ digit class) while later layers control finer aspects of speech (e.g.\ speaker identity).

Concretely, we can project two utterances $X_1$ and $X_2$ to their latent representations $\mathbf{w}_1$ and $\mathbf{w}_2$.
Then -- recalling the architecture in Fig.~\ref{fig:1_model_arch} -- we can use different $\mathbf{w}$ vectors as input into each \texttt{Style Block}.
According to our design motivation of the anti-aliasing filters in Sec.~\ref{subsec:anti_alias_filter}, the \textit{coarse styles} are captured in the earlier layers, and the \textit{fine styles} are introduced in later layers.
So we can perform voice conversion from $X_1$'s speaker to $X_2$'s speaker by
conditioning later layers with $\mathbf{w}_2$ while still using $\mathbf{w}_1$ in earlier layers.
This causes the generated utterance to inherent the speaking style (fine) from the target utterance $X_2$, but retain the word identity (coarse) from $X_1$.
By doing the opposite we can also do speech editing: having the speaker from $X_1$ say the word in $X_2$ by conditioning earlier layers with $\mathbf{w}_2$ while keeping $\mathbf{w}_1$ in later layers.
Furthermore, because the $W$ latent space is continuous, we can interpolate between retaining and replacing the coarse and fine styles to achieve varying degrees of voice conversion or speech editing.
We do these style mixing experiments in Sec.~\ref{subsec:voice_conversion}.
In preliminary experiments we found it optimal to use the first 11 \texttt{Style Block}s for coarse styles and the remaining 5 \texttt{Style Block}s for fine styles.

\subsubsection{Speech enhancement}
Speech enhancement is the task of removing noise from an %
utterance~\cite{speech_enhancement_benesty2006speech}.
Intuitively, if \modelname{}'s $W$-space is linearly disentangled, there should be a single direction corresponding to increasing or decreasing the background noise in an utterance.
Given
several utterances only varying in degrees of noise, we can project them to the $W$-space and compute the direction in which to move 
to change the
noise level.
Concretely, to denoise an utterance $X_0$, we can generate $N$ additional utterances by
adding increasingly
more Gaussian noise, %
providing %
a list of %
utterances $X_0, X_1, ..., X_{N}$, with $X_n = X_0 + \mathcal{N}(\mathbf{0}, n \sigma^2 \mathbf{I})$. %
We then project each utterance to the latent space, yielding $\mathbf{w}_0, \mathbf{w}_1, ..., \mathbf{w}_N$.
To get a single vector
corresponding to the direction of decreasing noise,  we compute
the average unit vector from %
the higher-noise vectors to the original latent vector: %
$$ \boldsymbol{\delta} = \frac{1}{N} \sum_{n=1}^N \frac{\mathbf{w}_0 - \mathbf{w}_n}{||\mathbf{w}_0 - \mathbf{w}_n||_2} $$
Now we can denoise the original utterance %
$X_0$ by %
moving in the direction of $\boldsymbol{\delta}$ in the latent space.
In Sec.~\ref{sec:8_unseen} we evaluate this method and investigate %
how far we can move in the $\boldsymbol{\delta}$-direction. %

\subsubsection{Speaker verification and keyword classification} \label{subsec:spk_verification_method}
The previous downstream tasks were generative in nature. %
The
$W$ latent space also allows 
us to perform
discriminative tasks such as speaker verification and keyword classification: %
determining which speaker or word is present %
in a given utterance, respectively. 
For both tasks we use the linear nature of disentanglement: given enrollment utterances containing labelled speakers (for speaker verification) or words (for keyword classification), we invert them to their $\mathbf{w}$ vectors. 
Then we find the linear projection within the $W$ latent space that %
maximizes the separation of the labeled characteristic using linear discriminant analysis (LDA).
For inference on new data, we invert the input, %
project it along the LDA axes, %
and make %
a decision based on linear distances %
to other points in the LDA-projected latent space.
The idea is to perform a linear projection that maximizes separation based on the chosen characteristic (e.g.\ speaker identity), otherwise the linear distance could be influenced more by other factors (e.g. noise).

In speaker verification, we need to predict a score corresponding to whether two utterances are spoken by the same speaker~\cite{yang2021superb}.
The speakers are both unseen during training.
So, we project all enrollment and test utterances to the $W$ latent space  
and then to the LDA axes (with axes fit on training data to maximize speaker separation).
The speaker similarity score is then %
computed as the %
cosine distance between two vectors in this space.
For evaluation, these scores can be used to compute an
equal error rate (EER)~\cite{yang2021superb}.
Similarly, for keyword classification, we project everything to the LDA axes maximizing content (i.e. what digit is spoken), %
compute the centroids of points associated with each digit (since we know the word labels beforehand), and then assign new utterances and their corresponding projected points to the label of the closest centroid by cosine distance.
We compare our latent-space LDA-based speaker verification and keyword classification approaches to task-specific models in Sec.~\ref{sec:8_unseen}.

\section{Experimental Setup: Unseen Tasks} \label{sec:exp_setup_unseen}

\subsection{Evaluation metrics}

To evaluate \modelname{}'s performance in each unseen task, we use the standard objective metrics from the literature:

\begin{itemize}
    \item \textit{Voice conversion}:
    We measure %
    conversion intelligibility following~\cite{benji_softvc, 2021gen_modelling_from_audio}, %
    whereby we perform voice conversion and then apply a speech recognition system to the output and compute a character error rate (CER) and $F_1$ %
    classification score to the word spoken in the original utterance.
    Speaker similarity is measured as described in~\cite{benji_softvc} whereby we find similarity scores between real/generated utterance pairs using a trained speaker classifier, and then compute an EER 
    with real/generated scores assigned a label of 0 and real/real pair scores assigned a label of 1. 
    \item \textit{Speech enhancement}: Given a series of original clean and noisy utterances, and the models' denoised output, we compute standard measures of denoising performance: narrow-band perceptual evaluation of speech quality (PESQ)~\cite{pesq_metric} and short term objective intelligibility (STOI) scores~\cite{stoi_metric}.
    \item \textit{Speaker verification}: Using the similarity scores on randomly sampled pairs of utterances from matching and non-matching speakers, we compute an EER as the measure of performance.
    We pair each evaluation utterance with another utterance from the same or different speaker with equal probability.
    \item \textit{Keyword classification}: We use the standard classification metrics of accuracy and $F_1$ score.  %
\end{itemize}

\begin{figure*}[t!]
\centering
\centerline{\includegraphics[trim={1cm 0.4cm 1cm 2.3cm}, clip, width=1.0\linewidth]{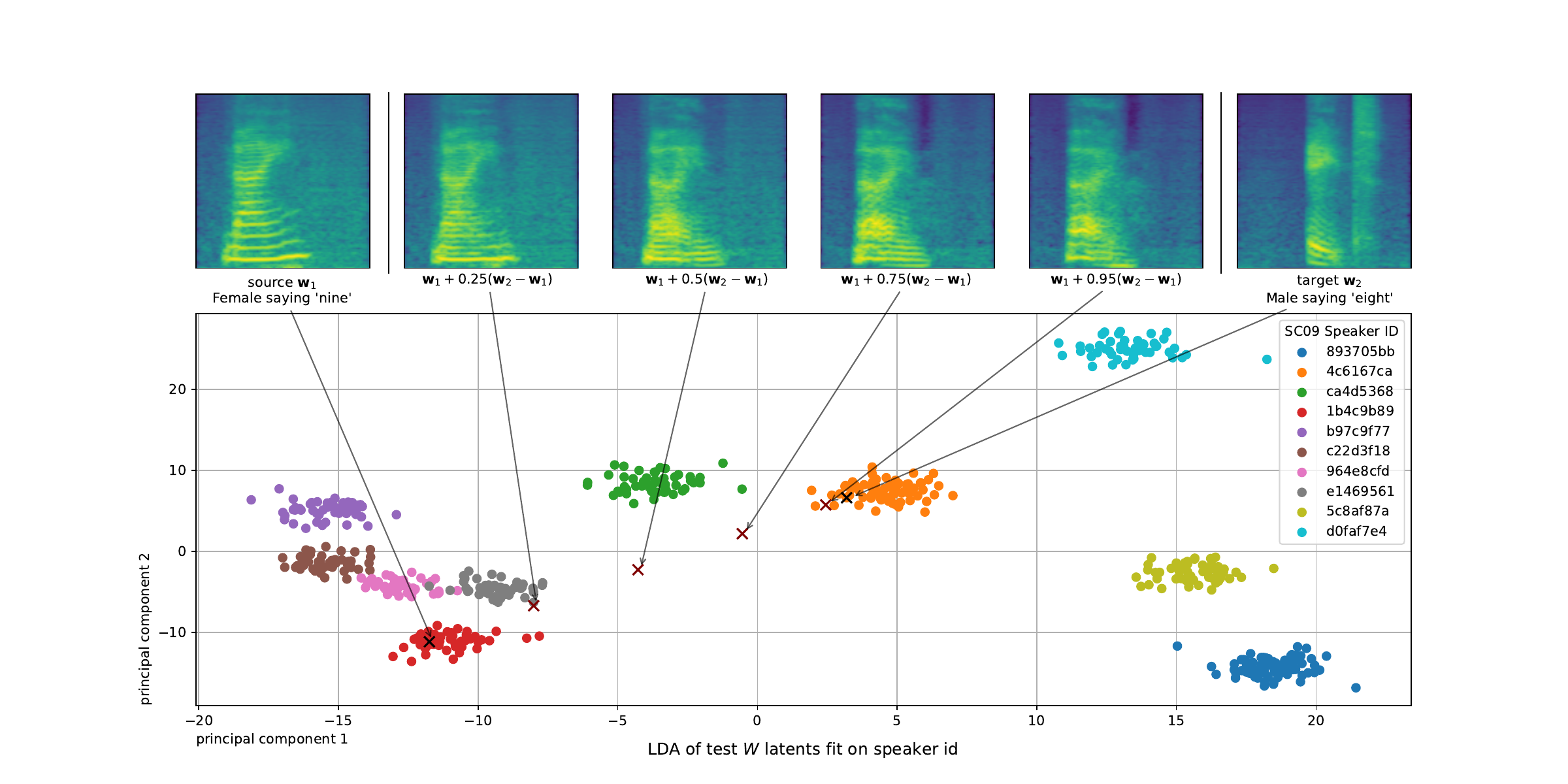}}
\caption{
    Voice conversion interpolation in the $W-$latent space.
    Given a source (top left) and target utterance (top right), we smoothly convert the speaker from the source to the reference by linearly interpolating the projected $\mathbf{w}$ vectors from $\mathbf{w}_1$ (source) to $\mathbf{w}_2$ (target) for use in the fine styles.
    The $W$-space interpolation is illustrated as a 2D linear discriminant analysis (LDA) decomposition of the SC09 test set.
}
\label{fig:2_vc_interp}
\end{figure*}

\subsection{Baseline systems} \label{subsec:5.B.2-uncond-baselines}

For each downstream task, we compare \modelname{} to a strong task-specific baseline trained on the SC09 dataset.
For voice conversion, we compare against AutoVC~\cite{autovc} -- a well-known voice conversion model.
Specifically, we use the model and training setup defined in \cite{autovc} and trained on the SC09 dataset\footnote{\scriptsize\url{https://github.com/RF5/simple-autovc}}.
For speech enhancement, we compare to MANNER~\cite{park2022manner}, a recent high-performing speech enhancement model operating in the time-domain. 
Since we have no paired clean/noisy utterances for the SC09 dataset, we follow the technique from~\cite{kashyap21_construct_noisy_ds} to construct a speech enhancement dataset.
Concretely, we assume each utterance in the SC09 dataset is a clean utterance and randomly add additional Gaussian noise to each %
utterance corresponding to a signal-to-noise ratio of 0 to %
10~dB, sampled uniformly.
We then train MANNER using the settings and code provided by the original authors\footnote{\scriptsize\url{https://github.com/winddori2002/MANNER}} on this constructed dataset.
For speaker verification, we use an x-vector model from~\cite{GE2E}, trained using the code and optimization settings from an open-source implementation\footnote{\scriptsize\url{https://github.com/RF5/simple-speaker-embedding}} on the SC09 training set, taking the final checkpoint with best validation performance.
Finally, for keyword %
classification we, use the same ResNeXT classifier described in Sec.~\ref{subsec:5-uncond-metrics}.

\section{Results: Unseen Tasks} \label{sec:8_unseen}

We apply \modelname{} to a range of tasks that it didn't see during training, comparing it to established baselines. 
Note, however, that the goal isn't to outperform these tailored systems on every task,
but rather to show that a single model, \modelname{}, can provide robust performance across a range of tasks that it was never trained on. 

\subsection{Voice conversion} \label{subsec:voice_conversion}

Following Sec.~\ref{subsec:style_mix_explanation}, we use \modelname{} for voice conversion.
We compare its performance to that of AutoVC (Sec.~\ref{subsec:5.B.2-uncond-baselines}).
To test the models, we sample
reference utterance from a different target speaker for each utterance in the SC09 test set, yielding 4107 utterance pairs on which we perform inference.
Through initial validation experiments, we found that the optimal interpolation amount for the \textit{fine styles} was to set the input to the later \texttt{Style Blocks} as $\mathbf{w}_1 + 1.75(\mathbf{w}_2 - \mathbf{w}_1)$ for $\mathbf{w}_1$ from the source utterance and $\mathbf{w}_2$ from the reference.

The results are shown in Table~\ref{tab:4_generative}.
In terms of similarity to the target speaker, %
AutoVC %
is superior to \modelname{}, while \modelname{} is superior in terms of %
intelligibility. %
Recall that our goal is not to outperform specialized models trained for specific tasks, but to demonstrate that \modelname{} can generalize to diverse tasks unseen during training.
While not superior to AutoVC, the scores in Table~\ref{tab:4_generative} are competitive in voice conversion. %

To give better intuition to the linear nature of the latent space, we show an example of interpolating the \textit{fine styles} smoothly from one speaker to another in Fig.~\ref{fig:2_vc_interp}: we project
the latent $\mathbf{w}$ points from many utterances in the test set using LDA fit on the speaker label.
The figure raises two interesting observations. First, we see that speaker identity is largely a linear subspace in the latent space because of the strong LDA clustering observed.
Second, as we interpolate from one point to another in the latent space, the intermediary points are still intelligible and depict realistic mixing of the source and target speakers, while leaving the content unchanged.
We encourage the reader to listen to audio samples at {\url{https://rf5.github.io/slt2022-asgan-demo}}.

\setlength{\tabcolsep}{3.6pt}
\begin{table}[tb]
    \renewcommand{\arraystretch}{1.2}
    \centering
    \caption{
        Unseen generative task performance  of \modelname{} compared to task-specific systems (AutoVC~\cite{autovc} for voice conversion, MANNER~\cite{park2022manner} for speech enhancement).
        } %
    \tablecaptionsep
    \label{tab:4_generative}
	\begin{tabularx}{0.99\linewidth}{@{}
            L
            S[table-format=2.1]
            S[table-format=2.1]
            S[table-format=2.1]
            S[table-format=1.2]
            S[table-format=2.1]
            @{}}
		\toprule
		& \multicolumn{3}{c}{{Voice conversion} (\%)} & \multicolumn{2}{c}{Speech enhancement}  \\
		\cmidrule(l){2-4}  \cmidrule(l){5-6}
		Model & {EER $\ \uparrow$} & {CER $\ \downarrow$} & {F1 $\ \uparrow$} & {PESQ$\ \uparrow$} & {STOI$\ \uparrow$}  \\
		\midrule
\textit{Task-specific baseline} & \ubold 32.3 & 64.8 & 29.1 & \ubold 2.13 & \ubold 81.4 \\
\modelname{} (HuBERT) & 29.7 & \ubold 37.6 & \ubold 61.3 & 0.95 & 21.2 \\
		\bottomrule
	\end{tabularx}
\end{table}
\setlength{\tabcolsep}{6pt}

\subsection{Speech enhancement}

\begin{figure*}[t!]
\centering
\centerline{\includegraphics[trim={1cm 0.4cm 1cm 2.3cm}, clip, width=1.0\linewidth]{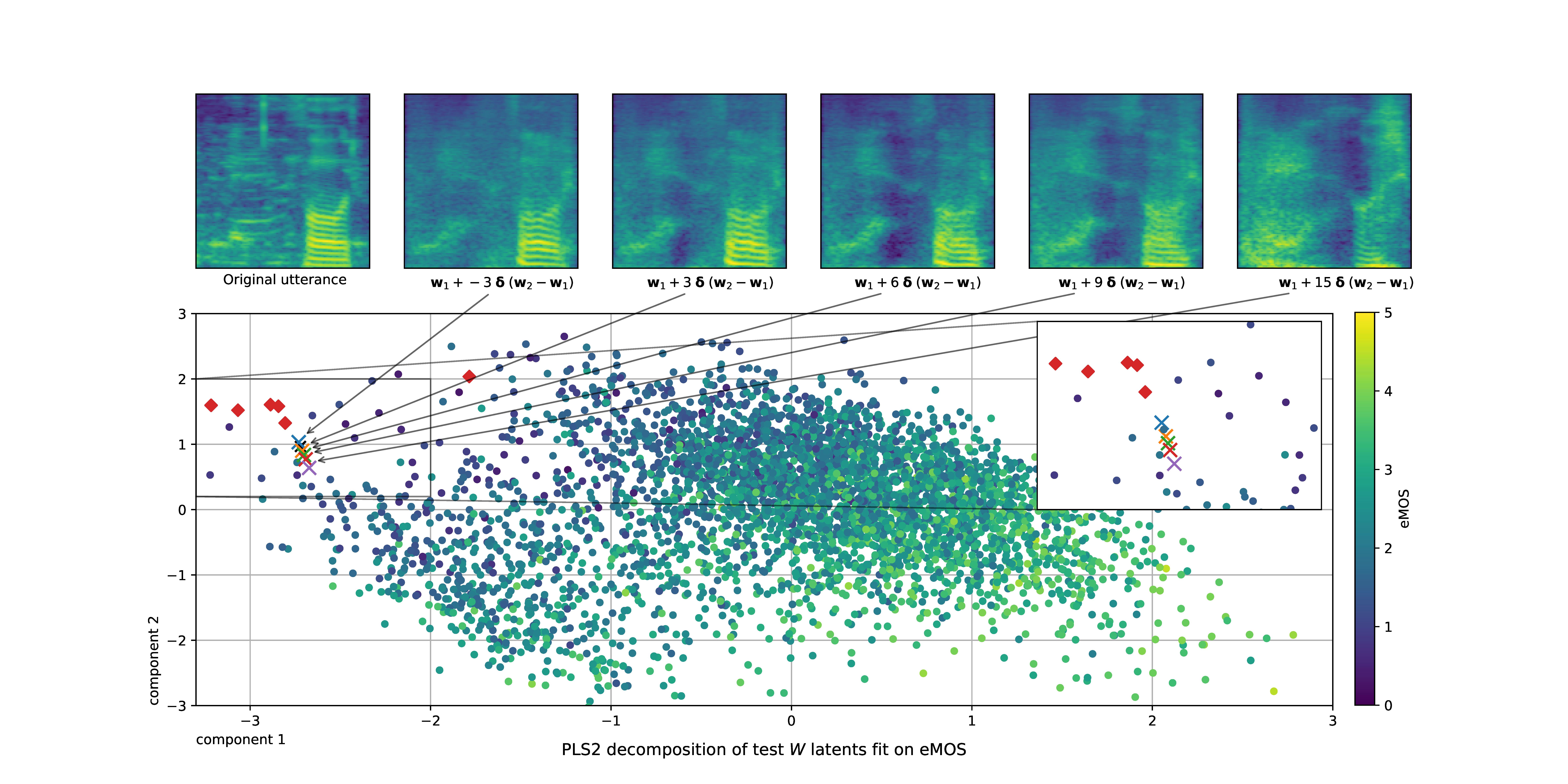}}
\caption{
    Speech enhancement in the $W$-latent space.
    Given an input utterance (top left) we add noise to it several times and find average direction of decreasing noise.
    Denoising or increasing noise is then performed by traversing in the latent space by varying amounts in the denoising direction $\boldsymbol{\delta}$ (top).
    The $W$-space interpolation is illustrated as a 2D partial least-squares regression (PLS2) of all the latent points from the SC09 test set fit on eMOS value as a indication of noise level.
    Latent variables corresponding to added noisy utterances used to estimate $\boldsymbol{\delta}$ are shown in red ({\footnotesize $\color{red} \blacklozenge$}).
}
\label{fig:3_speech_enhance}
\end{figure*}

Following Sec.~\ref{subsec:style_mix_explanation}, we  use \modelname{} for speech enhancement, with $N=10$, $\sigma^2=0.001$.
Using the process described in Sec.~\ref{subsec:5.B.2-uncond-baselines}, we evaluate both \modelname{} and the task-specific %
MANNER system %
on the SC09 test set in Table~\ref{tab:4_generative}.
From the results %
we see that \modelname{} performs worse than MANNER. %
As was the case in voice conversion, this isn't unexpected, given that
MANNER is a specialized task-specific model,
trained 
specifically for speech enhancement.
Meanwhile, \modelname{} is only trained to accurately model the density of utterances in the SC09 dataset.
The performance of \modelname{} on this task indicates some level of denoising.
We suspect some of the performance limitations are %
due to the relatively simple inversion %
method of Sec.~\ref{subsec:projection}, which allows for only a %
crude approximation of the ideal vector $\mathbf{w}$ 
corresponding
to a speech utterance $X$.
Using a more sophisticated latent projection method such as pivot tuning inversion~\cite{pti_roich2021pivotal} would likely reduce the discrepancy between $\Tilde{X}$ and $X$.
Our goal here, however, was simply to illustrate \modelname{}'s denoising capability without any additional fine-tuning of \modelname{}'s generator.

In a qualitative analysis, the process of moving along the $\boldsymbol{\delta}$ direction (Sec.~\ref{sec:4_latent_ops}) is graphically illustrated in Fig.~\ref{fig:3_speech_enhance}, where traversing certain amounts corresponds to a denoising action.
The figure shows the original utterance together with spectrograms where we subtract multiples of $\boldsymbol{\delta}$ (ideally increasing noise) or add multiples of $\boldsymbol{\delta}$ (ideally decreasing noise).
Alongside this, we find a 2D decomposition using partial least-squares regression (PSL2) fit on eMOS of utterances in the test set as a proxy for amount of noise present.
The key aspect of Fig.~\ref{fig:3_speech_enhance} is that we can see our motivation validated:
the utterances constructed by adding more Gaussian noise 
(shown as {\footnotesize $\color{red} \blacklozenge$} in Fig.~\ref{fig:3_speech_enhance}) broadly lie in the direction of decreasing eMOS in the PSL2 projection.
In the 2D projection, we see that moving in the $\boldsymbol{\delta}$ direction roughly corresponds to an increase in eMOS (a proxy for decreasing noise).
And, from the associated spectrograms in Fig.~\ref{fig:3_speech_enhance}, we see that we can increase and decrease noise to a moderate degree without significantly distorting the content of the utterance. 
A caveat of this method is that we cannot keep moving in the $\boldsymbol{\delta}$ direction indefinitely -- beyond $9\boldsymbol{\delta}$ we start to see extreme distortions including changes to the digit and speaker identity (remember the 2D plot in Fig.~\ref{fig:3_speech_enhance} contains all SC09 test points).
So, for the metrics computed in Table~\ref{tab:4_generative} we fix the denoising operation as interpolating with $9\boldsymbol{\delta}$.

\subsection{Speaker verification}

\setlength{\tabcolsep}{3.6pt}
\begin{table}[tb]
    \renewcommand{\arraystretch}{1.2}
    \centering
    \caption{
        Unseen discriminative task performance of \modelname{} compared to task-specific systems (GE2E RNN~\cite{GE2E} for speaker verification, ResNeXT classifier~\cite{resnext_xie2017aggregated} for keyword classification).
        } %
    \tablecaptionsep
    \label{tab:5_discriminative}
	\begin{tabularx}{0.99\linewidth}{@{}
            L
            S[table-format=2.2]
            S[table-format=2.2]
            S[table-format=2.2]
            @{}}
		\toprule
		& {Speaker verification} & \multicolumn{2}{c}{Keyword classification}  \\
		\cmidrule(l){2-2}  \cmidrule(l){3-4}
		Model & {EER $\ \downarrow$} & {Accuracy $\ \uparrow$} & {F1 $\ \uparrow$}  \\
		\midrule
\textit{Task-specific baseline} & \ubold 8.43 & \ubold 98.20 & \ubold 98.20 \\
\modelname{} (HuBERT) & 41.54 & 72.58 & 72.78 \\
\modelname{} (HuBERT) + LDA & 30.96 & 90.82 & 90.82 \\
		\bottomrule
	\end{tabularx}
\end{table}
\setlength{\tabcolsep}{6pt}

\begin{figure*}[t!]
\centering
\centerline{\includegraphics[trim={1cm 0.4cm 1cm 2.3cm}, clip, width=1.0\linewidth]{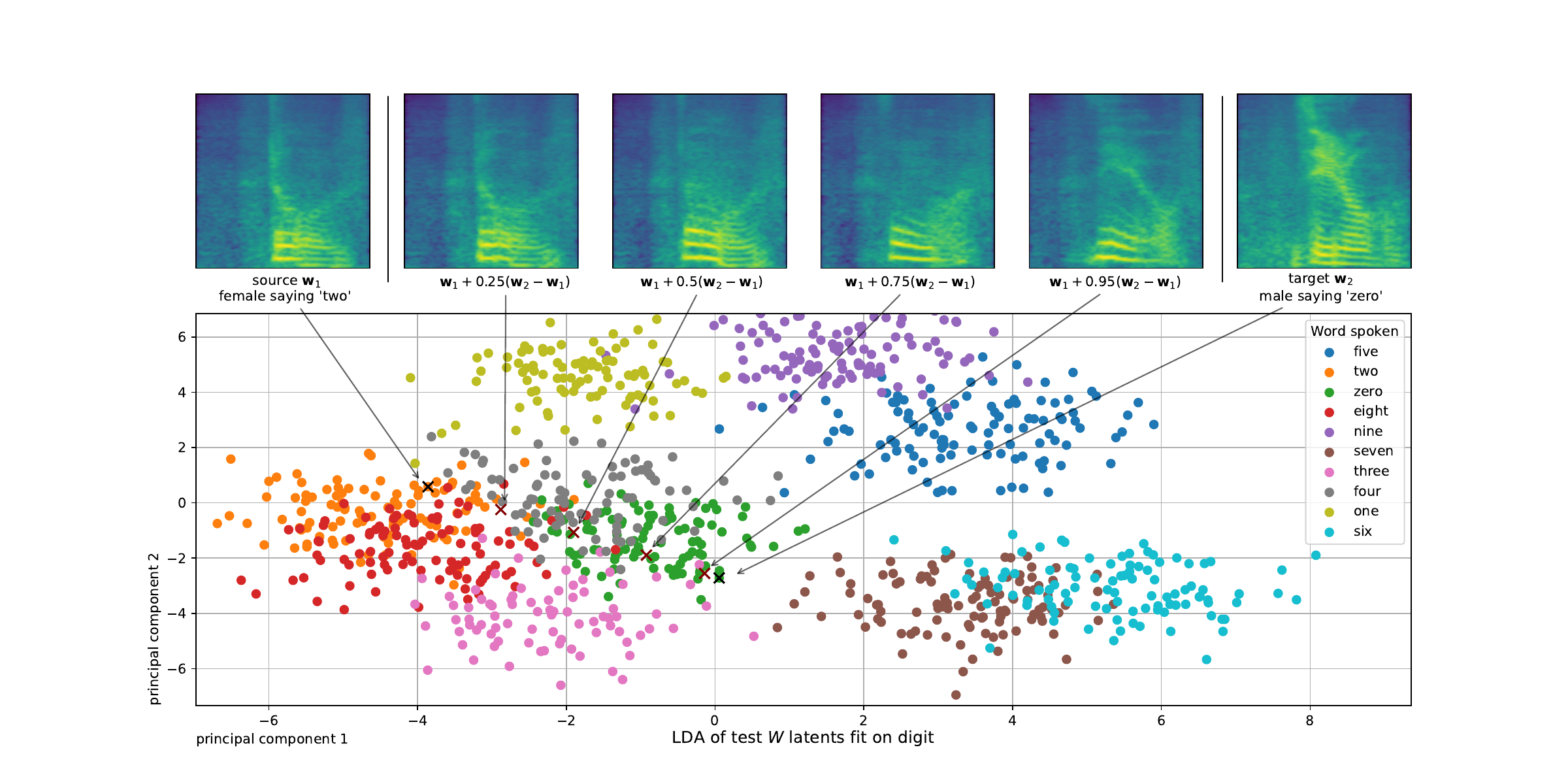}}
\caption{
    Keyword classification and speech editing in the $W$-latent space.
    Given a source (top left) and target utterance (top right), we smoothly convert the content from the source to the reference by linearly interpolating the projected $\mathbf{w}$ vectors from $\mathbf{w}_1$ (source) to $\mathbf{w}_2$ (target) for use in the coarse styles.
    The $W$-space interpolation is illustrated as a 2D LDA decomposition on the SC09 test set fit on spoken digits, showing its usefulness for keyword classification.
}
\label{fig:4_spk_classification}
\end{figure*}

We next consider a discriminative task: speaker verification, 
as outlined in
Sec.~\ref{subsec:spk_verification_method}.
Using the evaluation protocol
of Sec.~\ref{subsec:5-uncond-metrics},
we obtain the results
in Table~\ref{tab:5_discriminative} on the SC09 test set.
The first \modelname{} entry performs the naive cosine comparison, while the `+ LDA' entry compares distances after first applying LDA as described in Sec.~\ref{subsec:spk_verification_method}.
As with the generative tasks above, we see that \modelname{} is competitive, but not superior to the task-specific GE2E model~\cite{GE2E}.
The naive comparison in the regular latent space achieves fairly poor results.
However,
comparing the naive comparison method to the LDA projection approach,
we see an over 10\% EER improvement, with \modelname{} achieving %
an EER of 30.96\% despite never seeing
any of the speakers in the test set.  
This reinforces the observations from Fig.~\ref{fig:2_vc_interp}, where we can see that the LDA projection of latent $\mathbf{w}$ vectors %
disentangles speaker identity from other aspects of speech.
And,
because LDA is a linear operation,
we have not discarded any content information.
The speaker verification performance could likely be improved if a deep model was trained to cluster %
speakers in the latent space.
But, again, our goal is not state-of-the-art performance on this task, but to show that \modelname{} can achieve reasonable accuracy on a task for which it was not trained---owing to its
linearly disentangled latent space.

\subsection{Keyword classification}

Finally we apply \modelname{} to keyword classification.
We fit LDA on digit spoken using the SC09 validation set as enrollment, and compute predictions following Sec.~\ref{subsec:spk_verification_method}.
The results compared to the ResNeXT classifier are given in Table~\ref{tab:5_discriminative} using both the naive and LDA comparison method.
As before, we see a large performance jump when using the LDA projection,
with the best \modelname{} method only performing roughly 7\% worse than the specialized classification model.
A graphical illustration of how %
word classes are disentangled through LDA
is given in Fig.~\ref{fig:4_spk_classification}.
We
show
both the LDA projection of latent points (colored by their ground truth labels) and also %
how this disentanglement can be used to perform speech editing using the \textit{fine styles} described in Sec.~\ref{subsec:style_mix_explanation}, where we replace the word spoken but leave the speaker identity and other characteristics of the utterance intact. %
It allows us to smoothly interpolate
between one spoken word and another. This could potentially
be useful, e.g., for perceptual speech experiments. %
From both the LDA plot in Fig.~\ref{fig:4_spk_classification} and Table~\ref{tab:5_discriminative}, we again observe strong evidence that our design %
for
disentanglement is successful.

\section{Conclusion} \label{sec:9_conclusion}

We introduced \modelname{}, a model for unconditional speech synthesis designed to learn a disentangled latent space.
We adapted existing and incorporated new GAN design and training techniques to enable \modelname{} to learn a continuous, linearly disentangled latent space in order to outperform existing autoregressive and diffusion models. 
Experiments on the SC09 dataset demonstrated that \modelname{} outperforms previous state-of-the-art models on most unconditional speech synthesis metrics, while also being substantially faster.
Further experiments also demonstrated the benefit of the disentangled latent space: \modelname{} can, without any additional training, perform several speech processing tasks in a zero-shot fashion through linear operations in its latent space, showing reasonable performance in
voice conversion, speech enhancement, speaker verification, and keyword classification.

One major limitation of our work is scale: %
once trained, \modelname{} can only generate utterances of a fixed length, and the model struggles to generate coherent full sentences on datasets with longer utterances (a limitation shared by existing unconditional synthesis models~\cite{adv_audio_synth_donahue2018adversarial,diffwave_kong2020,sashimi_goel2022s}).
Furthermore,
the method used to project utterances to the latent space
could be improved by incorporating recent inversion methods~\cite{pti_roich2021pivotal}.
Future work will aim to address these shortcomings.

\bibliographystyle{IEEEtran}
\bibliography{IEEEabrv,references}

\end{document}